\documentclass[a4paper]{PoS}
\usepackage{amsmath}
\usepackage{graphicx}
\usepackage{caption}
\title{Constraining photon dispersion relations from observations of the Vela pulsar with H.E.S.S}

\ShortTitle{Constraining photon dispersions relations from observations of the Vela pulsar with H.E.S.S}

\author{\speaker{Mathieu Chr\'etien}\\
        LPNHE, Universit\'e Pierre et Marie Curie Paris 6, Universit\'e Denis Diderot Paris 7, CNRS/IN2P3, 4 Place Jussieu, F-75252, Paris Cedex 5, France.\\
        E-mail: \email{chretien@lpnhe.in2p3.fr}}

\author{Julien Bolmont\\
        LPNHE, Universit\'e Pierre et Marie Curie Paris 6, Universit\'e Denis Diderot Paris 7, CNRS/IN2P3, 4 Place Jussieu, F-75252, Paris Cedex 5, France.\\
        E-mail: \email{bolmont@lpnhe.in2p3.fr}}
    
\author{Agnieszka Jacholkowska \\
    	LPNHE, Universit\'e Pierre et Marie Curie Paris 6, Universit\'e Denis Diderot Paris 7, CNRS/IN2P3, 4 Place Jussieu, F-75252, Paris Cedex 5, France.\\
    	E-mail: \email{agnieszka.jacholkowska@lpnhe.in2p3.fr}} 

\author{For the H.E.S.S. Collaboration } 
    
\abstract{Some approaches to Quantum Gravity (QG) predict a modification of photon dispersion relations due to a breaking of Lorentz invariance. The effect is expected to affect photons near an effective QG energy scale. This scale has been constrained by observing gamma rays emitted from variable astrophysical sources such as gamma-ray bursts and flaring active galactic nuclei. Pulsars exhibit a periodic emission of possibly ms time scale. In 2014, the H.E.S.S. experiment reported the detection down to 20 GeV of gamma rays from the Vela pulsar having a periodicity of 89 ms. Using a likelihood analysis, calibrated with a dedicated Monte-Carlo procedure, we obtain the first limit on QG energy scale with the Vela pulsar. In this paper, the method and calibration procedure in use will be described and the results will be discussed.}

\FullConference{The 34th International Cosmic Ray Conference,\\
		30 July- 6 August, 2015\\
		The Hague, The Netherlands}

\begin{document}
\section{Introduction}
Two leading scenarios (String Theory (ST) and Loop Quantum Gravity (LQG)) compete for a theoretical description of Quantum Gravity (QG) (for a review see, e.g., \cite{Smolin}). They both predict a natural energy scale E$_{QG}$ at which gravitational and quantum effects are of the same order of magnitude. This energy is expected to be close to the Planck energy scale E$_p$ = 1.22 $\times$ 10$^{19}$ GeV or lower.
Several classes of QG models predict a violation of Lorentz invariance (LIV). This breaking could be the result of e.g., the foamy structure of the space-time in ST or due to a discrete and fluctuating space-time in LQP. This has become a very important window \mbox{on QG phenomenology \cite{Amelino1}.}
In presence of LIV, speed of photons is expected to depend on their energy $E$:
\begin{equation}
c'\approx c \times \left[1 \pm \frac{n+1}{2}\left(\frac{E}{E_{QG}}\right)^n\right],\text{    } n=
\begin{cases}
1 & \text{linear correction,}\\ 2 & \text{quadratic correction}
\end{cases}
\end{equation}
The photon propagation could be "superluminal" ($+$ sign) or "subluminal" ($-$ sign).
As a result two photons of energies $E_1$ and $E_2$ ($E_2>E_1$) emitted at the same time from a source at distance $d$ would be received with a relative delay $\Delta t$. The ratio of the delay over the energy difference is expressed as follows: 
\begin{equation}
\frac{\Delta t}{E_2^n-E_1^n} \simeq \pm \frac{(1+n)}{2} \frac{d}{c}\frac{1}{E_{QG}^n}, \label{fom}
\end{equation}
where $d$ is the euclidean distance of the source.
This expression allows time of flight measurements using high energy $\gamma$ rays emitted from astrophysical variable sources such as active galactic nuclei (AGN), gamma-ray bursts (GRB) or pulsars.
First LIV results with the Crab pulsar were obtained by the VERITAS collaboration \cite{CrabZitzer}, and more recently by MAGIC \cite{LIVMagicCrab}. The observation of the Vela pulsar by the High Energy Stereoscopic System (H.E.S.S.) offers another pulsating probe of LIV.
\section{H.E.S.S. data and Vela pulsar from March 2013 to April 2014}
From March 2013 to April 2014, the 28-meter H.E.S.S. telescope collected 24 hours of good quality data from the Vela pulsar.
About 10000 pulsed events were recorded above $\sim$20 GeV at low zenith angle $<$ 40$^{\circ}$. The detection was confirmed using two independent monoscopic analysis pipelines.
The H-test \cite{Htest} gives a significance of 14.6$\sigma$ for a H value of 280.6. Defining the ON and OFF phase regions to [0.5,0.6] and [0.7,1] respectively, yields a Li$\&$Ma \cite{LiMa} significance of 12.8$\sigma$. More details about the Vela pulsar analysis can be found in \cite{VelaICRC}.
For LIV studies, the reconstructed energy range was restrained to 20-100 GeV, reducing the excess statistics to 9322 events with a signal-to-noise ratio of $\sim$ 0.025. 
\section{The maximum likelihood method}
The procedure initially proposed in \cite{Martinez} has been slightly modified to be more suited to the case of periodic sources.
Assuming two $\gamma$ rays emitted from a pulsar with a given energy difference $\Delta E$, a linear correction due to LIV would lead to a phase lag $\Delta \Phi\simeq \Delta t\times f_0$, where $f_0$ is the pulsar rotational frequency. Using relation (\ref{fom}) for $n$ = 1, the linear "phase lag parameter" is defined by:
\begin{equation}
\varphi_l\equiv \frac{\Delta \Phi}{\Delta E}= \frac{\Delta t}{\Delta E} \times f_0 = \pm \frac{d}{c}\frac{f_0}{E_{QG}} . \label{fomphase}
\end{equation}
The probability density function (pdf) of observing a photon at a rotational phase $\Phi$ and with energy $E$, is defined as follows: $P(E,\Phi;\varphi_l)=\omega_s \times P_s(E,\Phi;\varphi_l)+(1-\omega_s) \times P_b(E,\Phi)$. The first term describes the pulsed signal component which depends on the phase lag parameter. It is given by:
\begin{equation}
P_s(E,\Phi;\varphi_l)=C\int_{0}^{\infty} A_{eff}(E_{\star})\Lambda_s(E_{\star})R(E-E_{\star},\sigma(E_{\star}))F_s\bigl(\Phi-\varphi_lE_{\star}\bigr) dE_{\star} \label{LIVModel}
\end{equation} 
where $\Lambda_s(E_{\star})$ is the spectral distribution of the excess. $A_{eff}(E_{\star})$ and $R(E-E_{\star},\sigma(E_{\star}))$ are the acceptance of the 28-meter H.E.S.S. telescope and the energy response function (taking into account energy reconstruction bias and dispersion). Although these two ingredients depend on the zenith angle, their value is averaged over the collected dataset. $F_s$ is the template phasogram, namely the phase distribution that would be observed without LIV. The $C$ factor ensures the right normalization of the pdf in the domain of observables.
\begin{figure}
		\centering
		\includegraphics[width=0.6\textwidth]{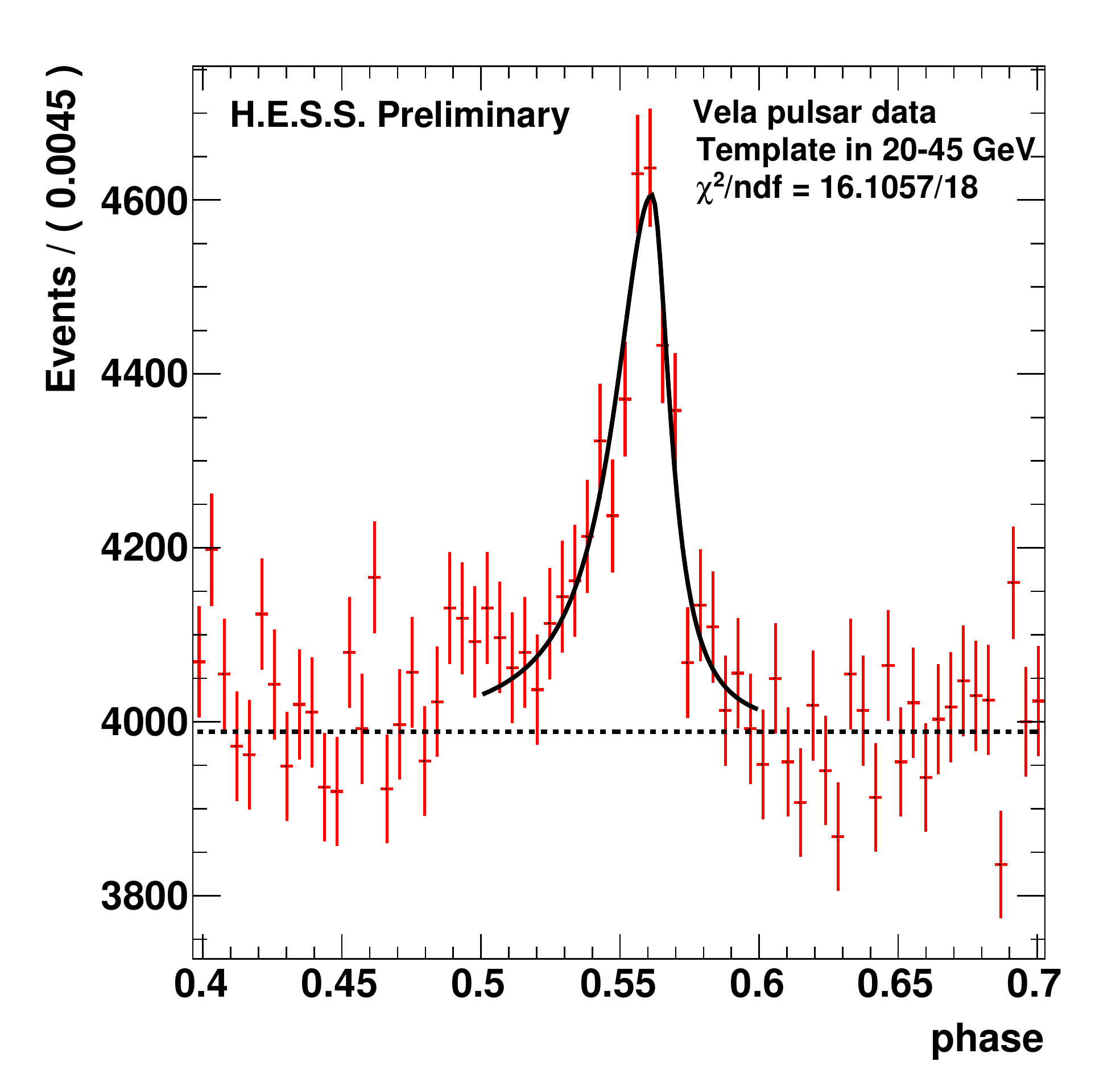}		
	\caption{Phasogram of the Vela pulsar in range 20-45 GeV for a bin width of 4.5$\times$10$^{-3}$ rotational phases. Data were collected from March 2013 to April 2014. The dashed line shows the background level determined in the OFF phase region. The solid line represents an asymmetrical Lorentzian fit to the pulsed events phase distribution. }
	\label{fig1}
\end{figure}
The term $P_b$ comprises mis-reconstructed hadrons, electrons, or diffuse $\gamma$ rays. The background contamination is expected to be uniformly distributed in phase and therefore not influenced due to LIV. The value $\omega_s$ is a relative weight between signal and background components.\\
The likelihood function is computed as the joint probability over all photons of a given dataset:
\begin{equation}
L(\varphi_l)=\prod_{i}P(E_i,\Phi_i;\varphi_l) \label{likelihood}.
\end{equation}
The minimum of $-2\Delta \ln(L)$ provides an estimate on $\varphi_l$ in units of rotational phase per TeV.\\
The background energy distribution in $P_b$ is parametrized with OFF phase events. $\Lambda_s(E_{\star})$ is obtained in ON phase region by fitting the excess distribution with a power law convoluted by the instrument response functions taken at the averaged zenith angle value of the data. The template phasogram is obtained with low energy events (20-45 GeV). The background level is first determined in the OFF phase region. Second the pulsed emission is fitted using an asymmetrical Lorentzian function plus a constant, set to the background level:
\begin{equation}
f(\Phi)=B+\begin{cases}
\frac{A}{1+\frac{(\Phi-\mu)^2}{\sigma_L^2}}, & \text{if }\Phi<\mu \\
\frac{A}{1+\frac{(\Phi-\mu)^2}{\sigma_R^2}}, & \text{if }\Phi\geq\mu \\
\end{cases}
\end{equation} 
where $A=617\pm 49$, $\mu=0.561\pm 0.001$, $\sigma_L=0.017\pm 0.003$ and $\sigma_R= 0.008\pm 0.002$ are the fitted amplitude, pulse position, left hand and right hand widths respectively. The template parametrization is shown in figure \ref{fig1}. This shape gives the best significance ($\chi^2/$ndf $\sim 16.1/18$) over simpler models (e.g. gaussian)  and is the one used in \cite{Fermi}.
\section{Calibration of the method and systematics}
\begin{figure}[b!]
	\begin{minipage}[t]{.5\textwidth}
		\centering
		\includegraphics[width=\textwidth]{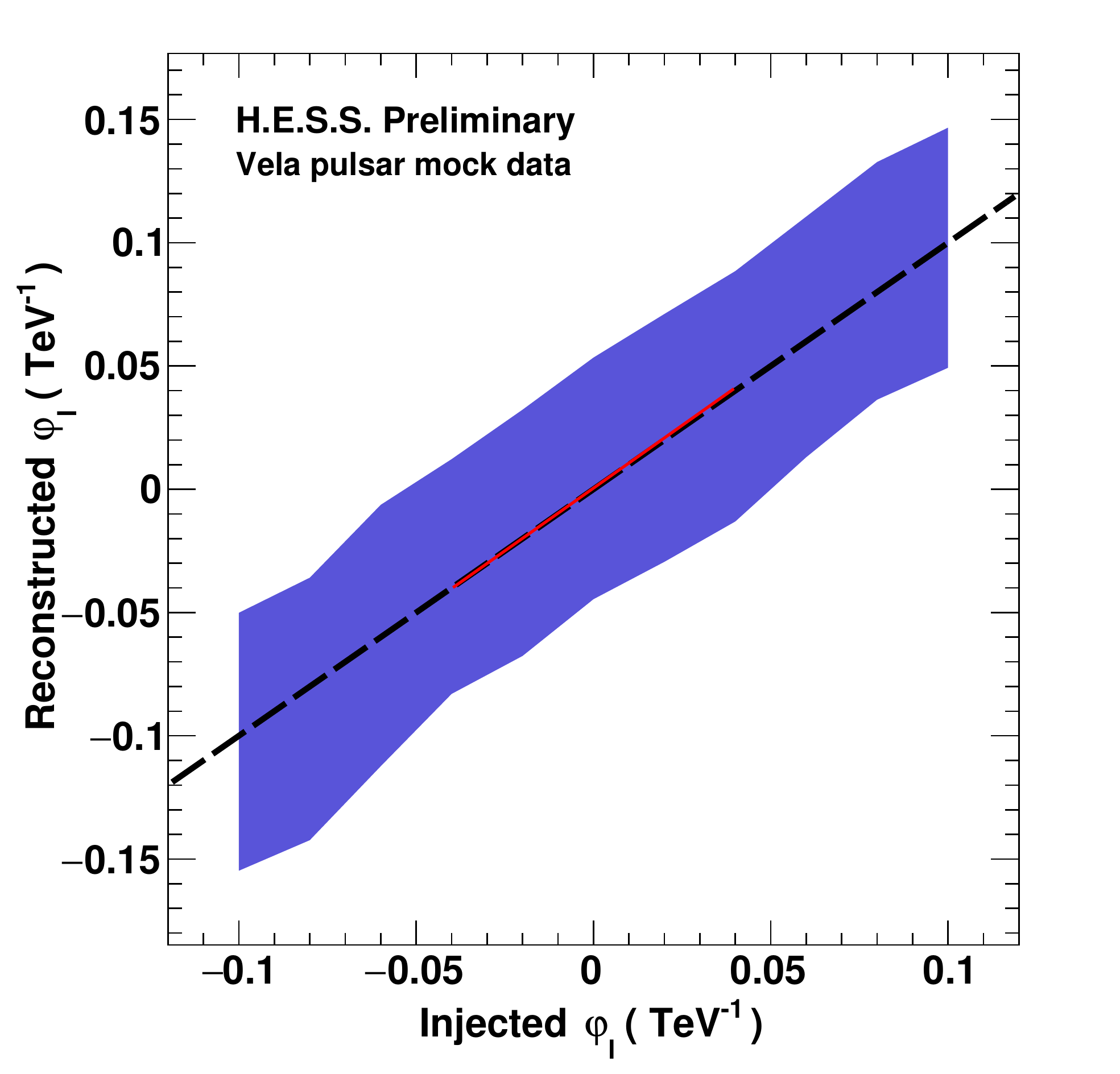}
	\end{minipage}\hfill	
	\begin{minipage}[t]{.5\textwidth}
		\centering
		\includegraphics[width=\textwidth]{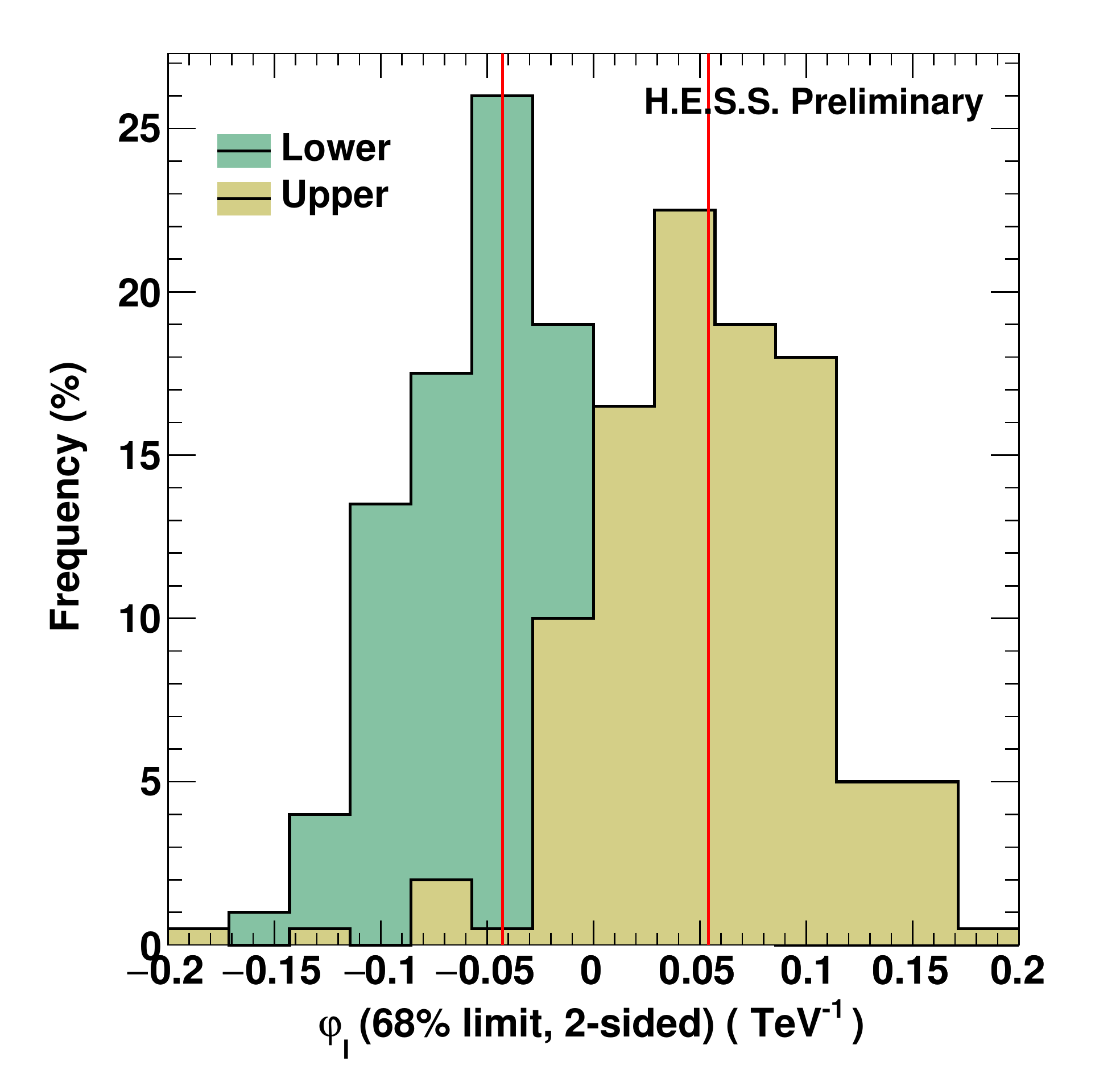}
	\end{minipage}\hfill		
	\caption{(Left) Reconstructed linear phase lag parameter $\bar{\varphi_l}$ as a function of the inject lag for Vela pulsar mock data. The contours are obtained from the dispersion of the reconstructed phase lag. The dashed line represents the perfect linear response. The red solid line is a linear fit to the curve. (Right) Distribution of the lower and upper bounds of the 68$\%$ (2-sided) CL reconstructed intervals, in the vicinity of $\varphi_l$=0. The red vertical lines stand for the mean value of the distributions.}
	\label{fig2}
\end{figure}
A dedicated toy Monte Carlo simulation software was developed. Hundreds of mock Vela pulsar data were simulated, varying the injected phase lag from -0.1 TeV$^{-1}$ to 0.1 TeV$^{-1}$ with step of 0.02 TeV$^{-1}$. For each value, the distribution of the reconstructed parameters was fitted with a gaussian function. It provides both the mean $\bar{\varphi_l}$ and dispersion $\sigma_{\varphi_l}$. 
The reconstructed phase lag $\bar{\varphi_l}$ as a function of the injected lag (so-called calibration curve) is shown in \mbox{figure \ref{fig2} (left)}. The blue contour represents the 1$\sigma_{\varphi_l}$ error on the reconstructed parameter. The slope of the \mbox{curve $\sim$ 1} indicates an almost unbiased phase lag measurement.
Anticipating results on data, confidence intervals were calibrated for a \mbox{68$\%$ (2 sided)} confidence level (CL). This was done in the vicinity of a null injected phase lag (see figure \ref{fig2}, right). The corresponding statistical error is extracted $\sigma_{\varphi_l(stat)}^{68\%}$=5$\times$10$^{-2}$ TeV$^{-1}$.

The tool was also employed to evaluate the systematics. The background contribution was studied in detail by varying the excess statistics and signal-to-noise ratio. 
Errors associated to the template parametrization, spectral index and instrument response functions were also propagated through the whole toy Monte Carlo chain. The table 1 summarizes the effect of various systematic uncertainties on the calibrated 68$\%$ CL intervals. Note that zenith dispersion systematics are related to the use of an averaged zenith angle value in the model definition (\ref{LIVModel}). The overall systematics on linear phase lag parameter ($\sigma_{\varphi_l(sys)}^{68\%}$=3$\times$10$^{-2}$ TeV$^{-1}$) are obtained by summing contributions of each term in quadrature. 
\section{Results, conclusions and prospect}
\begin{table}[t!]
	\centering
	\begin{tabular}{|c|c|c|}
		\hline
		Source & \multicolumn{2}{c|}{$\Delta\varphi_{l,i}$ (10$^{-2}$ TeV$^{-1}$)} \\ of systematics & lower bound& upper bound\\
		\hline
		Spectral index & < 1 & < 0.4 \\
		$F_s$ parametrization & < 1 & < 0.6 \\
	    Calibration curve & < 0.2 & < 0.2 \\
	    Background & < 0.8 & < 0.3 \\
	    Energy resolution & < 0.6 & < 1 \\
	    Energy bias & < 0.3 & < 1 \\
	    Acceptance factors & < 1 & < 1\\
	    Zenith dispersion & < 2 & < 0.7\\
	    Energy reconstruction & < 1 & < 1\\ \hline	  	    
	    $\sqrt{\sum_{i=1}^{9}\Delta\varphi_{l,i}^2}$ & < 3 & < 3\\\hline
	\end{tabular}
	\label{tab1}
	\caption{Influence of systematic uncertainties on the reconstructed linear phase lag parameter. The parameter changes are shown for the lower and upper bounds of the 68$\%$ CL intervals. Adding quadratically contribution of each terms gives the overall systematics.}
\end{table}
	
\begin{figure}[t!]
	\centering
		\includegraphics[width=0.6\textwidth]{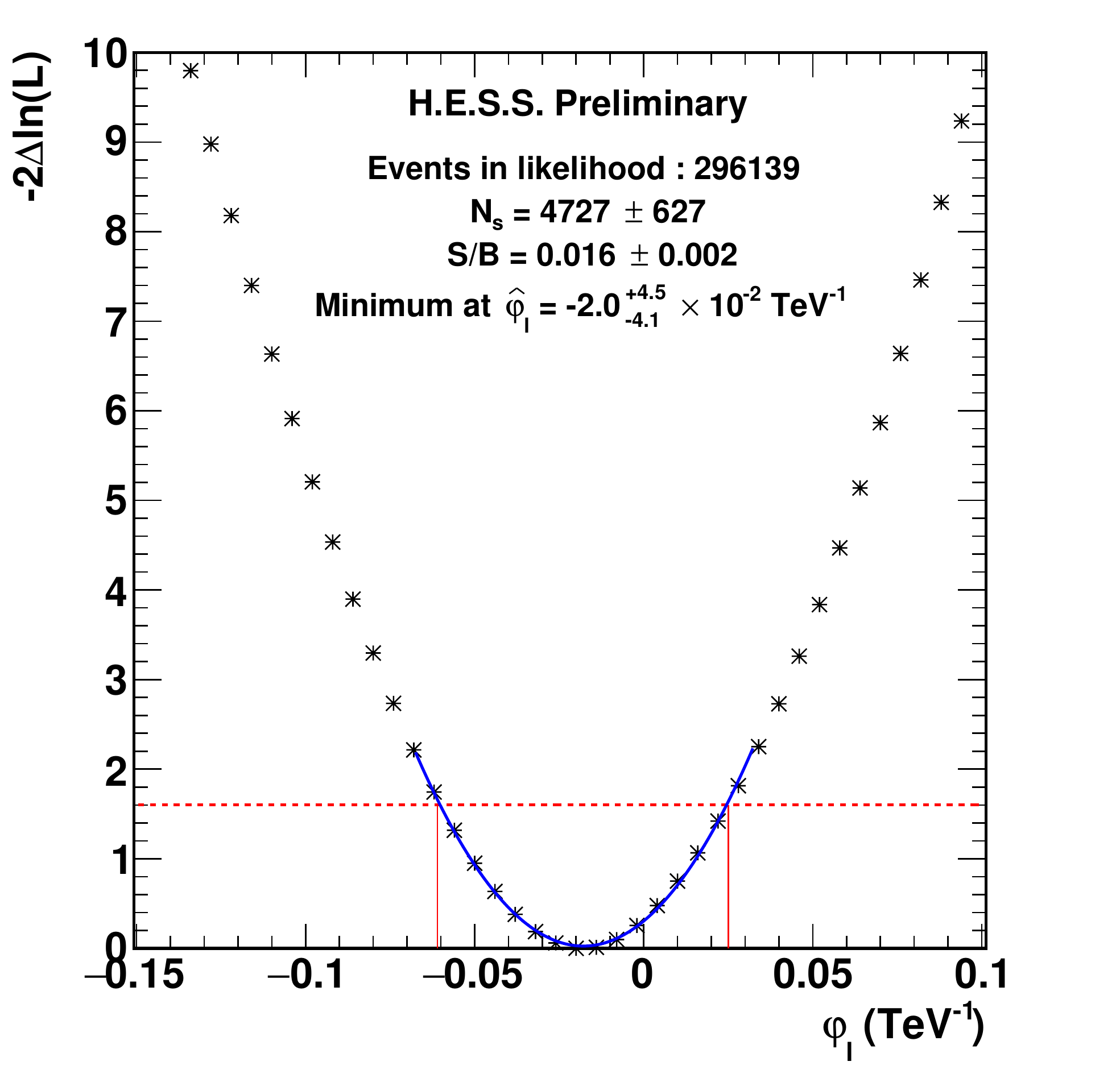}
	\caption{Curve of $-2\Delta\ln(L)$ as a function of $\varphi_l$ for a likelihood computation in 45-100 GeV. An estimate on $\varphi_l$ is given at the minimum location $\hat{\varphi_l}=-2.0^{+4.5}_{-4.1}\times 10^{-2}$ TeV$^{-1}$. The quoted 1$\sigma$ errors are obtained from the calibrated threshold for a 68$\%$ CL interval (red dashed line).}
	\label{fig3}
\end{figure}
An estimate on phase lag parameter is computed from data in energy range 45-100 GeV, using the parametrization as described in $\S$ 3. The curve of $-2\Delta\ln(L)$ is shown in figure \ref{fig3}. We derive a result on phase lag parameter from the location of the minimum, together with the 68$\%$ CL errors obtained from toy Monte Carlo studies in $\S$ 4:
\begin{equation}
\hat{\varphi_l}=\left(-2.0\pm 5.0_{(stat)}\pm 3.0_{(sys)}\right)\times 10^{-2} \text{  TeV$^{-1}$},
\end{equation}
No significant phase lag is found. Therefore the 95$\%$ CL lower limits on QG energy scale are derived summing statistical and systematic errors in quadrature: 
\begin{eqnarray}
E_{QG}^l &>& 3.72\times 10^{15} \text{  GeV,\space\space\space\space\space superluminal}\\
E_{QG}^l &>& 3.95\times 10^{15} \text{  GeV,\space\space\space\space\space     subluminal} \label{limit}
\end{eqnarray}
These constraints are more than one order of magnitude below the limits obtained with the Crab pulsar by VERITAS \cite{CrabZitzer} ($\sim$2$\times$10$^{17}$ GeV)  and MAGIC  \cite{LIVMagicCrab} ($\sim$4$\times$10$^{17}$ GeV for subluminal case). This is easily understood as the Crab pulsar is one order of magnitude farther than Vela and it is rotating faster by a factor $\sim$ 3. In addition and contrary to Vela, pulsed $\gamma$ ray emission was detected above 120 GeV from the Crab. The limits (\ref{limit}) are also four orders of magnitude below the most stringent limits obtained with \textit{Fermi}'s GRB \cite{GRBFermi}. For a review of the different existing limits \mbox{see e.g. \cite{Amelino1}.} \\
The relation (\ref{fom}) needs to be checked for any distances, therefore galactic pulsars, if they emit very high energy $\gamma$ rays similarly to the Crab are excellent candidates to probe the low distance regime. 
Due to a linearly increasing excess statistics with time, long term observation of Vela with H.E.S.S. would provide together with other pulsar observations an unique opportunity to improve sensitivity on $E_{QG}$. Considering 240 hours of live time observation, the sensitivity to LIV with the maximum likelihood method has been extrapolated, leading to a rough limit $E_{QG}$ $\gtrsim$1$\times$10$^{17}$ GeV. Although it is still far from the Planck scale it demonstrates the overall picture of the H.E.S.S. potential, searching for LIV with pulsars.
\newpage
\acknowledgments
The support of the Namibian authorities and of the University of Namibia in facilitating the construction and operation of H.E.S.S. is gratefully acknowledged, as is the support  of  the  German  Ministry of Education and  Research (BMBF), the Max Planck Society, the French Ministry of Research, the CNRS-IN2P3 and the Astroparticle Interdisciplinary Programme of the CNRS, the U.K. Particle Physics and Astronomy Research Council (PPARC), the IPNP of the  Charles University, the  South  African Department of Science and Technology and National Research Foundation, and by the University of Namibia.
We appreciate the excellent work of the technical support staff in Berlin, Durham, Hamburg, Heidelberg, Palaiseau, Paris, Saclay, and in Namibia in the construction and operation of the equipment. This work has been done within the Labex ILP (reference ANR-10-LABX-63) part of the Idex SUPER, and received financial state aid managed by the Agence Nationale 
de la Recherche, as part of the programme "\textit{Investissements d'avenir}" under the reference ANR-11-IDEX-0004-02.

\end{document}